\documentclass[fleqn,usenatbib]{mnras}

\usepackage{newtxtext,newtxmath}
\usepackage{enumerate}
\usepackage{gensymb}
\usepackage{textcomp}
\usepackage{graphicx}
\usepackage{float}

\usepackage[T1]{fontenc}
\usepackage{ae,aecompl}

\title[Global EoR signal with the Moon and the MWA]{Measuring the global 21-cm signal with the MWA-I: improved measurements of the Galactic synchrotron background using lunar occultation}

\author[B. McKinley et al.]
{B.~McKinley,$^{1,2,3}$\thanks{E-mail:ben.mckinley@curtin.edu.au}
G.~Bernardi,$^{4,5,6}$
C.~M.~Trott,$^{1,2,3}$
J.~L.~B.~Line,$^{7,3}$
R.~B.~Wayth,$^{1,2,3}$
\newauthor
A.~R.~Offringa,$^{8}$
B.~Pindor,$^{7,2,3}$
C.~H.~Jordan,$^{1,2,3}$
M.~Sokolowski,$^{1,2}$
S.~J.~Tingay,$^{1}$
\newauthor
E.~Lenc,$^{9,2}$
N.~Hurley-Walker,$^{1}$
J.~D.~Bowman,$^{10}$
F.~Briggs,$^{11}$
R.~L.~Webster$^{7,2,3}$
\\
$^{1}$International Centre for Radio Astronomy Research, Curtin University, Bentley, WA 6102, Australia\\
$^{2}$ARC Centre of Excellence for All-sky Astrophysics (CAASTRO),Curtin University, Bentley, WA 6102, Australia\\
$^{3}$ARC Centre of Excellence for All Sky Astrophysics in 3 Dimensions (ASTRO 3D), Curtin University, Bentley 6845 Australia\\
$^{4}$INAF-Istituto di Radioastronomia, via Gobetti 101, 40129, Bologna, Italy \\
$^{5}$Department of Physics and Electronics, Rhodes University, P.O. Box 94, Grahamstown, 6140, South Africa \\
$^{6}$Square Kilometre Array South Africa (SKA SA), Park Road, Pinelands 7405, South Africa \\
$^{7}$School of Physics, The University of Melbourne, Parkville, VIC 3010, Australia\\
$^{8}$Netherlands Institute for Radio Astronomy (ASTRON), Postbus 2, 7990 AA Dwingeloo, The Netherlands\\
$^{9}$Sydney Institute for Astronomy, School of Physics, The University of Sydney, NSW 2006, Australia\\
$^{10}$School of Earth and Space Exploration, Arizona State University, Tempe, AZ 8528USA\\
$^{11}$Research School of Astronomy and Astrophysics, Australian National University, Canberra, ACT 2611, Australia\\
}

\date{Accepted XXX. Received YYY; in original form ZZZ}

\pubyear{2017}

\hypersetup{draft}

\begin{document}
\label{firstpage}
\pagerange{\pageref{firstpage}--\pageref{lastpage}}
\maketitle

\begin{abstract}
We present early results from a project to measure the sky-averaged (global), redshifted $21\,$cm signal from the Epoch of Reionisation (EoR), using the Murchison Widefield Array (MWA) telescope. Because interferometers are not sensitive to a spatially-invariant global average, they cannot be used to detect this signal using standard techniques. However, lunar occultation of the radio sky imprints a spatial structure on the global signal, allowing us to measure the average brightness temperature of the patch of sky immediately surrounding the Moon. In this paper we present one night of Moon observations with the MWA between 72 -- 230~MHz and verify our techniques to extract the background sky temperature from measurements of the Moon's flux density. We improve upon previous work using the lunar occultation technique by using a more sophisticated model for reflected `earthshine' and by employing image differencing to remove imaging artefacts. We leave the Moon's (constant) radio brightness temperature as a free parameter in our fit to the data and as a result, measure $T_{\rm{moon}} = 180 \pm 12 $~K and a Galactic synchrotron spectral index of $-2.64\pm0.14$, at the position of the Moon. Finally, we evaluate the prospects of the lunar occultation technique for a global EoR detection and map out a way forward for future work with the MWA.

\end{abstract}




\section{Introduction}
\label{sec:intro}
Many experiments are now underway to detect and characterise redshifted $21\,$cm emission from the Epoch of Reionisation (EoR) and the cosmic dark ages. This signal provides the most promising avenue for exploring these early epochs in the Universe's history, however it is difficult to observe for two fundamental reasons. Firstly, the signal is weak and shrouded by bright astrophysical foregrounds and secondly, systematic instrumental effects contaminate the observations, requiring experiments to have unprecedented levels of calibration precision. 

Experiments to measure the redshifted $21\,$cm signal from the early Universe can be separated into two broad categories; those attempting to measure spatial fluctuations in the signal as a function of angular scale and frequency, and those attempting to measure the total power of the signal as a function of frequency, by averaging across the whole sky. The former category, with instruments such as the Murchison Widefield Array (MWA; \citealt{tingay2013,bowman2013,beardsley2016}), the LOw Frequency ARray (LOFAR; \citealt{lofar,patil2017}), the Precision Array to Probe the Epoch of Reionization (PAPER; \citealt{parsons2014}) and the Hydrogen Epoch of Reionization Array (HERA; \citealt{pober2014,deboer2016}), are employing radio interferometers to measure spatial fluctuations by means of power spectra. Future instruments, such as the Square Kilometre Array (SKA; \citealt{dewdney2010}), will be able to directly image the fluctuations \citep{koopmans2015}. 

The latter category of experiments, including the Experiment to Detect the Global EoR Signature (EDGES; \citealt{rogersbowman2008,rogers2008,bowman2010,monsalve2017b_constraints,bowman2018}), the Shaped Antenna measurement of the background RAdio Spectrum (SARAS;
\citealt{patra2013} and SARAS2; \citealt{singh2017a,singh2017b}), the Large Aperture Experiment to Detect the Dark Age (LEDA; \citealt{bernardi2016}), the Dark Ages Radio Explorer (DARE; \citealt{burns2012}), Discovering the Sky at the Longest wavelength (DSL; \citealt{DSL}), the Broadband Instrument for Global HydrOgen ReioNisation Signal (BIGHORNS; \citealt{bighorns}) and SCI-HI \citep{voytek2014} use (or will use) single-antennas to measure the all-sky or `global' signal. This approach has the advantage of an increased signal-to-noise ratio (SNR), but is challenging due to systematic instrumental effects \citep{monsalve2017a_calibration,singh2017a,bighorns}. 

The first detection of the global, redshifted $21\,$cm signal has been claimed by EDGES \citep{bowman2018}, who detect an absorption trough at 78~MHz, which is thought to coincide with the early cooling and then reheating of the gas in the Universe during Cosmic Dawn. While the frequency of the absorption trough observed by \citet{bowman2018} is expected from theoretical predictions and modelling of the Cosmic Dawn \citep{pritchard2010,cohen2017}, the large amplitude and flattened profile of the absorption trough require new physics to explain \citep{barkana2018}. Our observing band, with a lower bound of 72~MHz, partially overlaps with the observed absorption trough, which is 19~MHz wide. Therefore, the MWA lunar occultation experiment provides an important means to verify the EDGES result, since it is subject to very different systematics.

The idea to use the Moon as a thermal reference source in an interferometric measurement of the global EoR signal was first presented by \citet{shaver1999}. This approach, which has been investigated further by \citet{mckinley2013} and \citet{vedantham2015}, combines the advantages of an interferometer, such as immunity from frequency-dependent receiver-noise bias and the ability to spatially isolate reflected RFI signals, with the increased SNR of a global experiment. Using interferometric techniques also reduces the effects of frequency-dependent antenna beam response that can cause artificial spectral structure in global signal observations \citep[e.g.,][]{vedantham2014,bernardi2015,mozdzen2016}.

In this paper we present new results from lunar observations with the MWA, which demonstrate our ability to measure the average temperature of a Moon-sized patch of the sky across a frequency range 72 -- 230~MHz. This is the first paper in a series that will describe the MWA lunar occultation experiment. We begin by reviewing the theory behind the lunar occultation technique in Section~\ref{sec:theory}. In Section~\ref{sec:observations} we describe our observations with the MWA, followed by the details of the data reduction and modelling in Section~\ref{sec:modelling}, which includes earthshine mitigation. In section \ref{sec:results}, we present our results and analysis. In Section~\ref{sec:discussion} we discuss our results in the context of other experiments and the prospects for detecting the global EoR signal using the lunar occultation technique. We conclude in Section~\ref{sec:conclusion} and outline the path forward for the MWA lunar occultation experiment.

\section{Theory}
\label{sec:theory}

Interferometers are not completely insensitive to the global average temperature of the sky, but the sensitivity drops off rapidly as a function of baseline length such that only the very shortest baselines (a few wavelengths) are sensitive to the signal (see \citealt{singh2015} and \citealt{vedantham2015}). \citet{presley2015} suggested that the sensitivity to the global signal by an interferometer is due to the spatial imprint of the primary beam shape causing an overlap with the `zero baseline' in $(u,v)$ space, and that this could be exploited by using very closely spaced antennas, with sufficiently large apertures. However \citet{singh2015} have shown that it is in fact not possible to build such antennas close enough together to get any overlap with the origin in the $(u,v)$ plane, without shadowing of the antennas occurring. Rather, interferometers are inherently sensitive to a global signal because, even for a uniform sky, the coherence function sampled by an interferometer does extend beyond the origin, and there are methods that can be used to extend this sensitivity out to longer baseline lengths \citep{singh2015}.

One such method to extend the sensitivity of an interferometer to the global signal out to longer baseline lengths is the lunar occultation technique (see \citealt{vedantham2015} for a full description). Using this technique the flux density of the Moon, $S_m(\nu)$, as a function of frequency $\nu$, can be measured (so long as the angular scales corresponding to the Moon's disk are adequately sampled) and converted to a brightness temperature, $\Delta T(\nu)$ in K, by:
\begin{equation} 
\Delta T(\nu)=\frac{10^{-26}c^{2}S_m(\nu)}{2k \Omega\nu^{2}}, 
\label{eqn:Tmeasured1}
\end{equation}
where $\nu$ is the frequency in Hz, $c$ is the speed of light in m~s$^{-1}$, $k$ is the Boltzmann constant in units of m$^{2}$~kg~s$^{-2}$~K$^{-1}$, $S_m(\nu)$ is the measured flux density of the Moon in Jy and $\Omega$ is the solid angle subtended by the Moon, which, during these observations, was $7.365\times10^{-5}$~sr. We calculate $\Omega$ in sr by first calculating the angular area of the Moon in square degrees using the value of the Moon's angular radius on 2015 September 26 as seen from the MRO, computed using \sc{pyephem} \rm \citep{pyephem} and then dividing this by the angular area of one sr expressed in square degrees (see \citealt{guthrie1947}).

Since the `zero baseline', average signal is missing in the interferometer measurements, the brightness temperature $\Delta T(\nu)$ in K, represents the difference between the total lunar brightness temperature, $T_{\rm{lunar}}(\nu)$, and the average brightness temperature of the occulted patch of sky, $T_{\rm{sky}}(\nu)$:
\begin{equation} 
\begin{split}
\Delta T(\nu) & = T_{\rm{lunar}}(\nu) - T_{\rm{sky}}(\nu) \\
              & = [T_{\rm{moon}} + T_{\rm{refl-Earth}}(\nu) + T_{\rm{refl-Gal}}(\nu)] \\ & \qquad \qquad - [T_{\rm{Gal}}(\nu) + T_{\rm{CMB}} + T_{\rm{EoR}}(\nu)],
\end{split}
\label{eqn:deltat}
\end{equation}
where $T_{\rm{moon}}$ is the intrinsic (thermal) temperature of the Moon, $T_{\rm{refl-Earth}}(\nu)$ is the additional lunar brightness temperature due to reflected emission from the Earth or `earthshine', $T_{\rm{refl-Gal}}(\nu)$ is the additional lunar brightness temperature due to reflected Galactic radio emission, $T_{\rm{Gal}}(\nu)$ is the mean temperature of the occulted patch of sky due to Galactic radio emission, $T_{\rm{CMB}}=2.725$~K \citep{mather1994} is the Cosmic Microwave Background (CMB) and $T_{\rm{EoR}}(\nu)$ is the EoR signal, the detection of which is the ultimate goal of this project. For this paper, however, we do not include the EoR signal in our modeling as it is expected to be at least two orders of magnitude weaker than the foregrounds \citep{furlanetto2006}, which are dominated by Galactic synchrotron emission. Instead, it is the foreground signal, $T_{\rm{Gal}}(\nu)$, that we attempt to measure in this work, in order to prove the technique and assess its utility for EoR detection.

Earthshine is known to contaminate the brightness temperature of the Moon (see \citealt{sullivan1978,sullivan1985,mckinley2013,vedantham2015}). Radar studies of the Moon by \citet{evans1969} find that the reflective properties of the Moon at radio wavelengths depend strongly on both the angle of incidence of the radiation and on wavelength. The reflected signal from Earth-bound transmitters has two components. Firstly, there is a quasi-specular component due to radiation with small angles of incidence. This component results in a point-like source in the centre of the disk, which increases in brightness with increasing wavelength (since the Moon becomes `smoother' as the wavelength increases). Secondly, there is a disk component due to diffuse reflection from the rough surface of the Moon. The second component results from radiation with larger angles of incidence and increases in brightness with \emph{decreasing} wavelength , since the Moon becomes rougher as wavelength decreases \citep{beckmann1965a,beckmann1965b,klemperer1965,evans1969}. Our methods for mitigating against contamination by earthshine are described in Section~\ref{sec:observations}.

In equation~\ref{eqn:deltat} we assume that $T_{\rm{moon}}$ is a constant over the frequency range observed in our experiment. At higher frequencies the observed brightness temperature of the Moon varies with the phase of the Moon due to heating of the surface, however this effect decreases with frequency as the longer wavelengths of radiation originate from increasing depths below the surface of the Moon \citep{krotikov1964a} and at metre wavelengths it is thought to be negligible \citep{baldwin1961}. The only published measurements within our frequency range, at 178~MHz from \citet{baldwin1961}, report a value of $T_{\rm{moon}}=233 \pm 8$ K. In our analysis we leave $T_{\rm{moon}}$ as a free parameter in our fitting and therefore obtain a new estimate for the value of $T_{\rm{moon}}$ at low radio frequencies. In Section~\ref{sec:Tmoon} we investigate the value of $T_{\rm{moon}}$ in more detail, comparing our data to previous results. 

The $T_{\rm{sky}}(\nu)$ that we measure using the lunar occultation technique is also dependent on the Moon's position in the sky at the time it is observed. Since the Moon moves relative the background sky at approximately $0.54 \degree \rm{h}^{-1}$, over the course of the $\sim5 \rm{h}$ of observations described in this paper it traverses about $2.7\degree$. For simplicity we assume that the average sky temperature does not vary considerably over $2.7\degree$. 

In order to eventually extract the global EoR signal from measurements of $T_{\rm{sky}}(\nu)$, we will need to remove foreground emission by taking advantage of the spectral smoothness of the foregrounds, possibly using Bayesian techniques such as those of \citep{bernardi2016}. It will be necessary to increase our SNR by averaging over many patches of occulted sky and therefore foreground removal will need to be conducted separately for each night of observations. Due to the very low global-signal SNR on the one epoch of observations described in this paper, we leave foreground removal as a subject to be covered in paper II, which will include multiple epochs.

\section{Observations}
\label{sec:observations}

The Moon moves across the sky relative to the background of Galactic and extra-Galactic emission at a rate of approximately $13\degree$ per day. In our dedicated observing campaign with the MWA (project ID G0017), we take advantage of this fact by taking observations over two nights, separated by at least one night, and taking the difference between the images. As first suggested by \citet{shaver1999}, this allows us to remove imaging artefacts without the need for cleaning. For the on-Moon observations, we choose nights when the Moon transits at a high elevation angle ($>60\degree$) and is well-separated from the Galactic plane ($> 50\degree$). This allows optimal conditions for calibration and imaging. The tracking uses the standard approach for the MWA, where observations are divided into intervals with constant beamformer settings (232~s, in this case), resulting in a `drift and shift' mode of operation, rather than continuous tracking of the source. We track the position of the Moon in this way, cycling through five centre frequencies before re-pointing the telescope if necessary. For our observing strategy to work well, observations must be taken in pairs such that there is both an on-Moon and off-Moon observation that has the same beamformer settings and the same Local Sidereal Time (LST).

In this paper we include data from just two nights to demonstrate the technique and to measure the spectrum of the Galactic synchrotron emission for this particular MWA pointing. Observations tracking the Moon were taken on 2015 September 26 and the corresponding, LST-locked, `off-Moon' observations were taken on 2015 September 29. Five contiguous bands using the MWA's 30.72~MHz instantaneous bandwidth were used, covering the range 72--230~MHz, while avoiding approximately 5~MHz around the ORBCOMM satellites operating frequency of 137~MHz. Data were taken with time and frequency resolutions of 1~s and 40~kHz, respectively. The total on-Moon integration time per frequency channel is $\sim$70~min.

\section{Data reduction and modelling}
\label{sec:modelling}

The $(u,v)$ data are flagged for radio-frequency interference (RFI) and converted to measurement sets using \sc{cotter} \rm \citep{offringa2015}. Both the on- and off-Moon measurement sets are phase shifted so that the phase centre of each LST-locked pair of observations is at the coordinates of the Moon's position during the observation. The data are then calibrated using \sc{calibrate}\rm, developed by \citet{offringa2016}, and a sky model generated from multi-frequency radio data with the Positional Update and Matching Algorithm (\sc{puma}\rm; \citealt{line2017}). Dirty images in instrumental polarisation are produced using \sc{wsclean} \rm \citep{wsclean}. We use uniform weighting to avoid contamination from large-scale structure and increase our angular resolution for characterisation of quasi-specular earthshine. Although using uniform weighting does reduce the MWA sensitivity, on the scales associated with the Moon it does not have a significant effect. Using natural weighting for the imaging was trialled and found not to have a large impact on our results, apart from increasing the size of the error bars in the final spectrum. 

We use an image size of $2048 \times 2048$ pixels, covering a $\sim$$17\degree$ field of view and having a bandwidth of 1.28~MHz. The synthesised beam size ranges from a full-width-half-maximum (FWHM) of approximately 6~arcmin at the lowest frequency to 2~arcmin at the highest frequency. To simplify the modelling process which is to follow, the same pixel size of $0.0085\degree \ ($0\farcm51$)$ is used across the full frequency range. The images are primary-beam corrected and converted to Stokes images using \sc{pbcorrect} \rm \citep{offringa2016} and the MWA primary beam model of \citet{marcin}. To check for poorly calibrated images, a simple root-mean-square (rms) threshold check is made on each image. If the measured rms of a patch of the image close to, but not including, the Moon is below a specified threshold then the LST-locked on-Moon and off-Moon Stokes~I image pairs are differenced and saved. A point spread function (PSF) image is also produced at each frequency for each observation. 

\begin{figure*}
\centering 
\includegraphics[clip,trim=0 0 0 0,width=1.0\textwidth,angle=0]{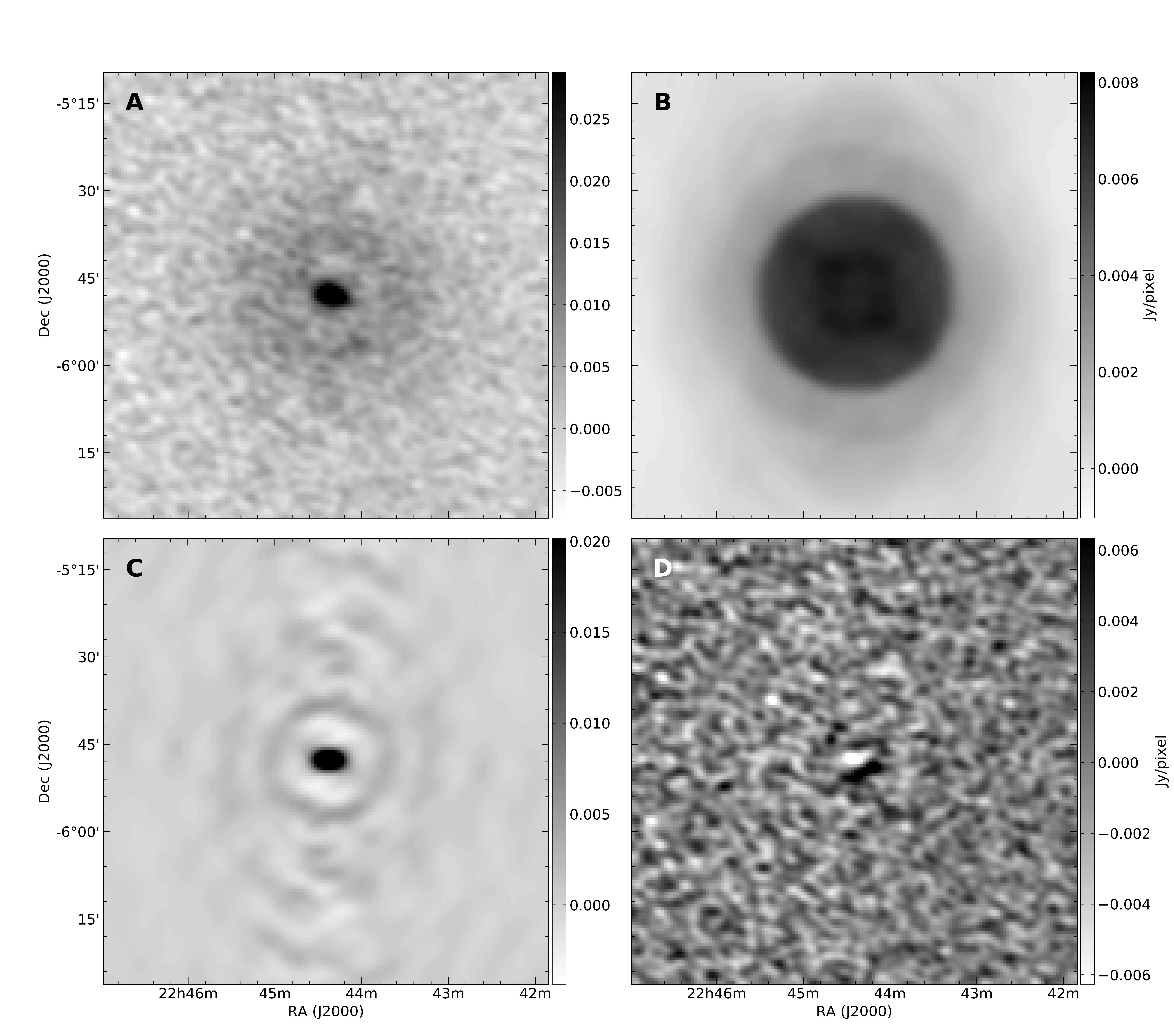}
\caption{Images resulting from the Moon modelling process, averaged over 16 observations, for the 1.28~MHz frequency channel centred on 200~MHz. Panel A: Average difference image. Panel B: Average reconstructed disk model of the Moon (convolved with the telescope PSF). Panel C: Average reconstructed quasi-specular earthshine (convolved with the telescope PSF). Panel D: Average residual image ($D=A-B-C$). All images are shown on the same angular scale.}
\label{modelling}
\end{figure*}

The ionosphere may have a strong impact on global EoR experiments, due primarily to frequency-dependent absorption and emission that can corrupt the redshifted $21\,$cm signal \citep{vedantham2014,datta2016}. However, \citet{iono} have shown that these effects may average to zero over long integrations, hence our approach to observe over many nights. We therefore do not consider these effects further in this paper. Refractive ionospheric effects, however, are particularly problematic for our experiment, which relies upon making difference images between two nights with potentially different ionospheric conditions. The ionospheric conditions over the MWA have been studied by \citet{jordan2017} who find that 74\% of MWA observations can be classified as having little to no active ionospheric activity. The data selected for this paper were chosen from such a quiet night, where ionospheric effects can effectively be ignored. This can be seen by the minimal mis-subtraction of radio sources in our difference images (see Fig.~\ref{modelling}, panels A and D). 

We implement a modelling procedure, building upon the work of \citealt{vedantham2015}, to estimate and isolate reflected earthshine and recover the flux density of the Moon's disk. \citet{vedantham2015} assume specular reflection for earthshine and model the Moon's total emission as a disk component and a point-source component due to earthshine. We introduce a more sophisticated model of the earthshine that more accurately captures the Moon's reflective properties. 

We employ a two-step earthshine-mitigation process. In the first step we follow \citet{vedantham2015}, but replace the point-source model of earthshine with a mask of $5\times8$ pixels ($2.5 \times 4$ arcmin) elongated along the RA axis. This is to account for two effects which broaden the quasi-specular component of the earthshine. The first is the angular broadening as described by \citet{vedantham2015}, who calculate that the broadening would have an rms width of $3.86'$ based on an rms slope for the lunar surface of $14\degree$, as estimated by \citet{daniels1963a}. The value used for the rms slope, however, is incorrect because \citet{daniels1963a} did not account for the diffuse component of the reflections in their analysis. Instead, we use the revised value of $6.8\degree$ for the rms slope of the Moon's surface \citep{daniels1963b}. The second broadening effect is due to the Moon's motion against the sky background over the course of each 4-min observation, which smears the quasi-specular component of the earthshine across the RA direction. Our model can be expressed as:
\begin{equation} 
{\bf D}=(s_{\rm{disk}}{\bf{M}} + s_{\rm{spec}}{\bf B})*{\bf{P}} + {\bf{N}},
\label{eqn:model}
\end{equation}
where ${\bf D}$ is the matrix of dirty difference image brightness values, ${\bf M}$ is a mask representing the disk of the Moon (having a value of 1 within the Moon's diameter and 0 elsewhere), ${\bf B}$ is the mask representing the broadened, quasi-specular component of the earthshine, ${\bf P}$ is the PSF of the telescope and ${\bf N}$ is noise. The variables $s_{\rm{disk}}$ and $s_{\rm{spec}}$ are the brightness values of the disk component and the quasi-specular reflection component of our Moon model, respectively, in units of Jy/pixel. 

We solve for $s_{\rm{disk}}$ and $s_{\rm{spec}}$ using the maximum likelihood estimation employed by \citet{vedantham2015}. The total integrated flux densities for the disk component, $S_{\rm{disk}}$, and the  quasi-specular component, $S_{\rm{spec}}$, are then calculated by multiplying $s_{\rm{disk}}$ and $s_{\rm{spec}}$ by their respective masks and summing over all pixels.

\begin{figure*}
\centering 
\includegraphics[clip,trim=0 25 20 40,width=1.0\textwidth,angle=0]{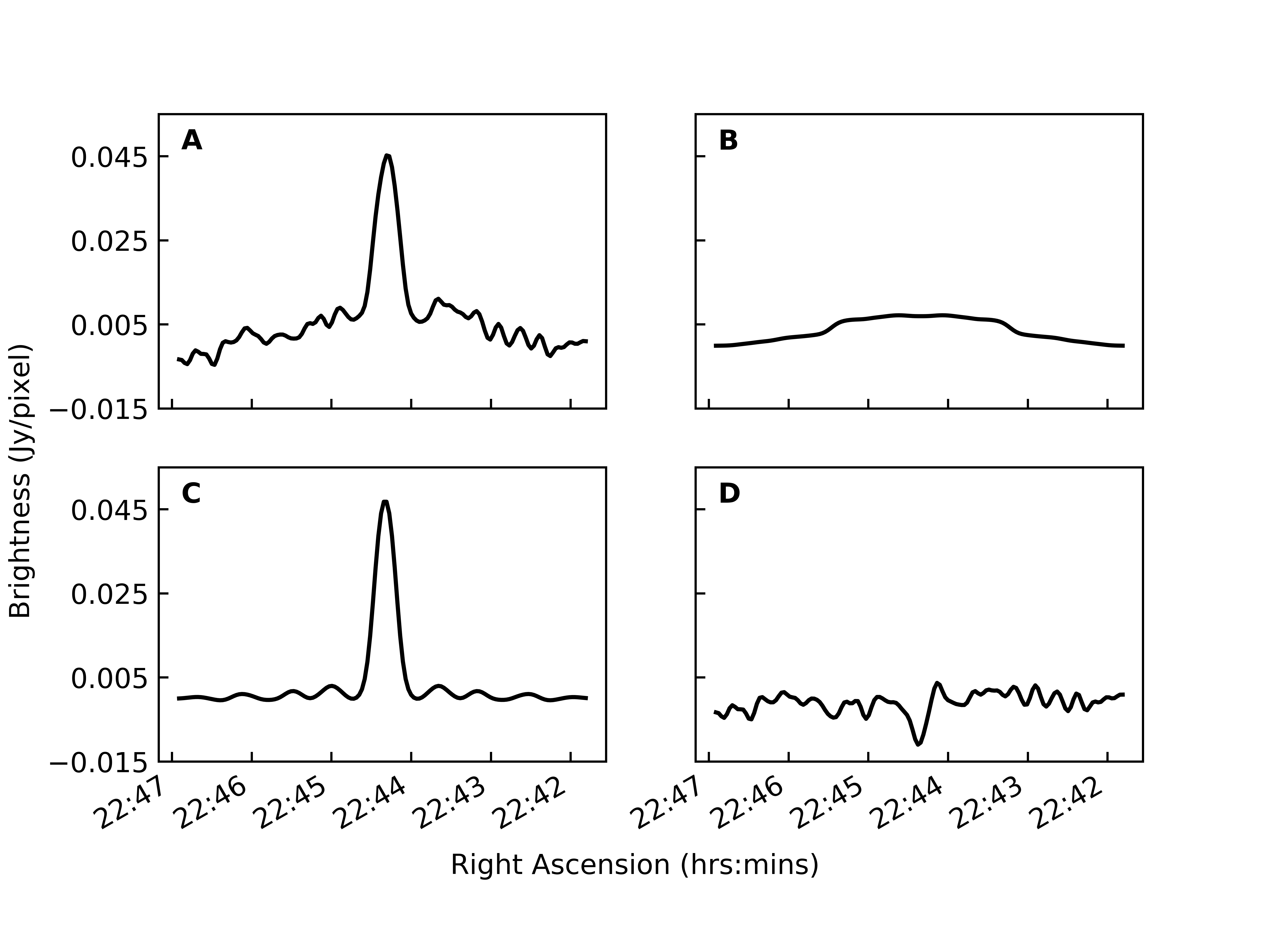}
\caption{Cross-sections through the centres of the images shown in Fig.~\ref{modelling}, along the RA axis. Panel A: Cross-section through the average difference image. Panel B: Cross-section through the average reconstructed disk model of the Moon (convolved with the telescope PSF). Panel C: Cross-section through the average reconstructed quasi-specular earthshine (convolved with the telescope PSF). Panel D: Cross-section through the average residual image. All plots are shown on the same y-axis brightness scale.}
\label{profiles}
\end{figure*}

In Fig.~\ref{modelling} we show the resulting images from the modelling process, averaged over 16 observations, for the 1.28-MHz frequency channel centred on 200~MHz. Panel A shows the average difference image, Panel B shows the average reconstructed disk component of the model (convolved with the telescope PSF), Panel C shows the average reconstructed point-like source of quasi-specular earthshine (convolved with the telescope PSF) and Panel D shows the average residual image ($D=A-B-C$). In Fig.~\ref{profiles} we show cross sections through the images in Fig~\ref{modelling}, along the RA axis, in order to give a clearer picture of how well the RFI mitigation is performing.

\begin{figure*}
\centering 
\includegraphics[clip,trim=0 10 0 0,width=1.0\textwidth,angle=0]{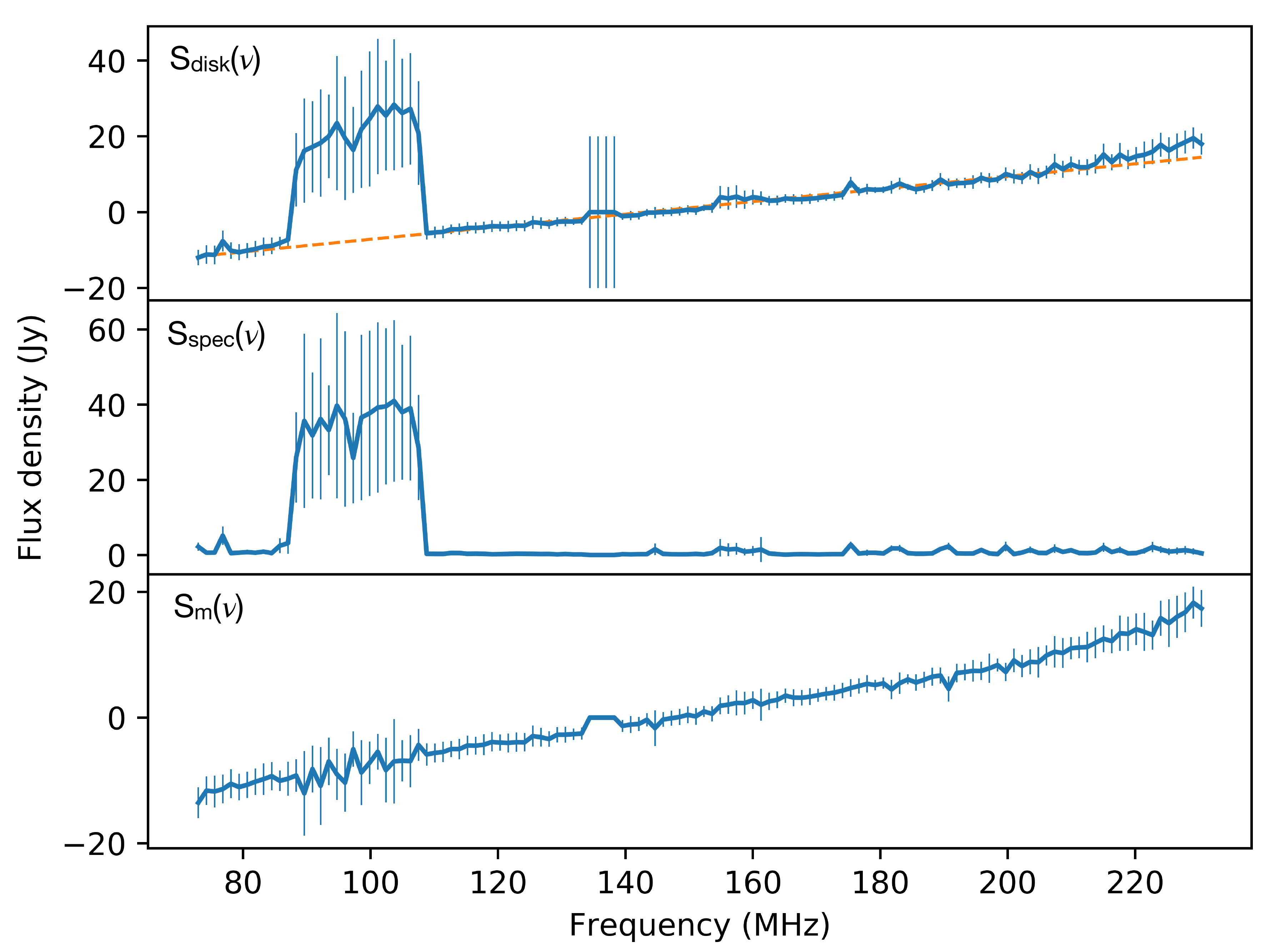}
\caption{Measurements of the integrated flux density for three components of our Moon and earthshine model. Data points are the mean of the measurements from multiple observations at each frequency and the error bars are the standard deviation. Top panel: Flux density of the disk component of our Moon model $S_{\rm{disk}}(\nu)$, resulting from the first stage of our earthshine-mitigation process. Centre panel: Flux density of the quasi-specular component of the earthshine $S_{\rm{spec}}(\nu)$, resulting from the first stage of our earthshine-mitigation process. Bottom panel: The final estimate of the intrinsic flux density of the Moon $S_{m}(\nu)$, after removal of the diffuse earthshine component in the second stage of our earthshine mitigation process, using equation \ref{eqn:final_Sm}. A colour version of this figure is available in the online article.}
\label{rfi_diffuse}
\end{figure*}

The modelling is performed on each observation separately and the mean and standard deviation of $S_{\rm{disk}}$ and $S_{\rm{spec}}$ are computed for each frequency band. The top and centre panels of Fig.~\ref{rfi_diffuse} show $S_{\rm{disk}}$ and $S_{\rm{spec}}$ as a function of frequency. The data points are the mean and the error bars are the standard deviation. We find that the standard deviations are much larger than the uncertainties computed in the modelling process and hence we use these in the plots and subsequent fits for the Galactic sky temperature and brightness temperature of the Moon. We postulate that these large standard deviations are due to the time variation of the reflected earthshine and the large channel bandwidth used, which would include many earthshine-contributing transmitters. These issues will be addressed in future work where we will model the earthshine contribution to the brightness temperature at higher temporal and spectral resolution.

A clear feature of both of these plots is the large spike of emission between 88-110~MHz. In both cases, this feature is due to emission from Earth in the FM radio band (see \citealt{mckinley2013}) being reflected by the Moon in both a quasi-specular (in the case of $S_{\rm{spec}}$) and diffuse (in the case of $S_{\rm{disk}}$) fashion. The large error bars in the region dominated by earthshine correspond to a large spread between values computed for the different observations. This is likely due to the time variability of the earthshine, which is expected to fluctuate over the course of the night as different transmitters on Earth move into and out of the line of sight of the Moon.

As discussed in Section~\ref{sec:theory}, the contaminating earthshine consists of two components; a quasi-specular component represented by $S_{\rm{spec}}$ and a diffuse component which we label $S_{\rm{diffuse}}$. The flux density of the Moon, $S_{m}(\nu)$, which we require to solve equation~\ref{eqn:Tmeasured1}, can therefore be expressed as:
\begin{equation} 
S_{m}(\nu) = S_{\rm{disk}}(\nu) - S_{\rm{diffuse}}(\nu).
\label{eqn:disk_components}
\end{equation}

In the second step of the earthshine mitigation process we aim to determine the ratio of the diffuse to quasi-specular earthshine components across the entire frequency band, $R_{\rm{e}}(\nu)$, so that we can remove the diffuse component of the earthshine from the disk flux density and obtain $S_{m}(\nu)$ using equation~\ref{eqn:disk_components}.

We combine equations~31 and 32 of \citet{evans1969} to compute the frequency dependence of the ratio between the diffuse and quasi-specular components of the earthshine, $R_{\rm{e}}(\nu)$ and obtain:
\begin{equation} 
R_{\rm{e}}(\nu) = \left(\frac{S_{\rm{diffuse}}(\nu)}{S_{\rm{spec}}(\nu)}\right)= A\nu^{0.58},
\label{eqn:ratio}
\end{equation}
where $S_{\rm{diffuse}}(\nu)$ is the flux density of the disk component of the earthshine, $S_{\rm{spec}}(\nu)$ is the flux density of the quasi-specular component of the earthshine, $\nu$ is frequency in MHz and A is a proportionality constant. 

Combining equations~\ref{eqn:disk_components} and \ref{eqn:ratio}, we can obtain $S_{m}(\nu)$ by:
\begin{equation} 
S_{m}(\nu) = S_{\rm{disk}}(\nu) - R_{\rm{e}}(\nu) S_{\rm{spec}}(\nu).
\label{eqn:final_Sm}
\end{equation}

The bright, reflected emission in the FM radio band is actually of benefit to us at this point, as it allows us to directly measure the proportionality constant A in equation~\ref{eqn:ratio} with a high SNR. We determine A by measuring $S_{\rm{diffuse}}(\nu)$ and $S_{\rm{specular}}(\nu)$ in the middle of the FM band at 100~MHz. We find $S_{\rm{diffuse}}(\nu=100)$ by first fitting a line to $S_{\rm{disk}}(\nu)$ to estimate the contribution of $S_{m}(\nu)$ and then subtracting the estimate of $S_{m}(\nu=100)$ from $S_{\rm{disk}}(\nu=100)$. We then calculate A by:
\begin{equation} 
A = \frac{S_{\rm{diffuse}}(\nu=100)}{S_{\rm{spec}}(\nu=100)}(100)^{-0.58}.
\label{eqn:A}
\end{equation}

At this point we can obtain $S_{m}(\nu)$ by applying equation~\ref{eqn:final_Sm} to the measured spectrum, however this is not particularly useful for 21-cm cosmology as it does not reduce the size of the error bars on $S_{m}(\nu)$. Instead, we re-run the modelling procedure and use our previously-computed values of $R_{\rm{e}}(\nu)$ to remove the diffuse component of the earthshine for each observation before calculating the mean and standard deviation at each frequency. In doing this we effectively remove the diffuse component of the RFI and the errors associated with its time variability. The final RFI-corrected measurements of $S_{m}(\nu)$ are shown in the bottom panel of Fig.~\ref{rfi_diffuse}.

\section{Results and Analysis}
\label{sec:results}
\subsection{Reflected Galactic Emission}
\label{sec:reflected}
Having measured the flux density of the Moon's disk and removed the contaminating earthshine emission, we can now proceed to determining the average background sky temperature using equation~\ref{eqn:deltat}. We must first, however, calculate the additional contribution to the Moon's brightness temperature due to the reflection of Galactic emission, $T_{\rm{refl-Gal}}(\nu)$.

\begin{figure*}
\centering 
\includegraphics[clip,trim=0 200 0 100,width=1.0\textwidth,angle=0]{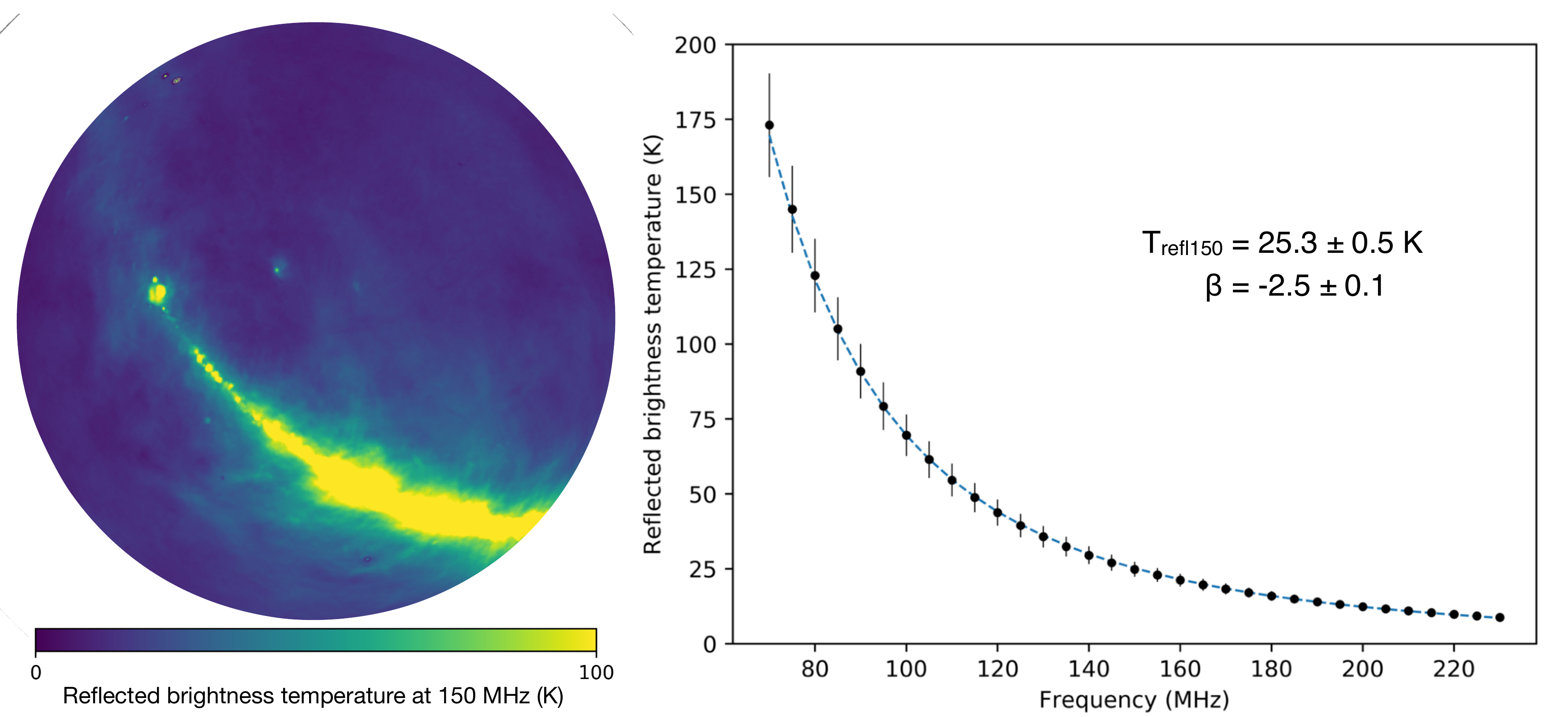}
\caption{Left panel: The reflected Galactic brightness temperature mapped onto the disk of the Moon as seen by the MWA at 150~MHz, computed using ray-tracing and assuming a Moon albedo of 7\%. The GSM of \citet{gsm} has been used as the sky model. In this map, the disk-average of the reflected temperature, $T_{\rm{refl150}}$, is 25.3~K. Right panel: The disk-averaged temperature calculated from the ray-tracing results, plotted as a function of frequency. The dotted line is a power law fit, assuming a 15\% error on each disk-averaged temperature computed using the GSM.}
\label{fig:reflected_galaxy}
\end{figure*}

At these frequencies the Moon has an albedo of 0.07 \citep{evans1969}, which means that there is a significant amount of Galactic radio emission that is reflected by the Moon's disk. We use ray tracing to calculate the expected reflected Galactic emission, using the positions and orientations of the Moon, the MWA and the Global Sky Model (GSM; \citealt{gsm}) at the time of the observations. The reflected Galactic emission mapped onto the disk of the Moon is computed at 5~MHz intervals from 70 to 230~MHz. The resulting map at 150~MHz is shown in Fig.~\ref{fig:reflected_galaxy} (left panel) as an example. Taking the disk-average of the reflected image, we find that the temperature due to reflection of Galactic emission ($T_{\rm{refl-Gal}}(\nu)$ from equation~\ref{eqn:deltat}) is well-modeled by a power-law:
\begin{equation} 
T_{\rm{refl-Gal}}(\nu)= T_{\rm{refl150}}\left(\frac{\nu}{150 \ \rm{MHz}}\right)^{\beta},
\label{eqn:reflected_gal}
\end{equation}
where $\nu$ is the frequency in MHz, $T_{\rm{refl150}}$ is the reflected temperature at 150~MHz taking into account the 7 per cent albedo of the Moon, found to be 25.3~K, and $\beta$ is the spectral index, found to be $-2.50$. The disk-averaged reflected temperatures, along with the power law fit, are shown in Fig.~\ref{fig:reflected_galaxy} (right panel).

\subsection{Fitting for $T_{\rm{Gal}}(\nu)$ and $T_{\rm{moon}}$ }
\label{sec:TGal}
Rearranging equation~\ref{eqn:deltat} and placing the unknown quantities on the left, we calculate:
\begin{equation} 
\begin{split}
T_{\rm{Gal}}(\nu) - T_{\rm{moon}} & = [T_{\rm{refl-Earth}}(\nu) + T_{\rm{refl-Gal}}(\nu)] \\ & \qquad \qquad - [\Delta T(\nu) + T_{\rm{CMB}} + T_{\rm{EoR}}(\nu)],
\end{split}
\label{eqn:TGal}
\end{equation}
using our measurements of the flux density of the Moon, $S_{m}(\nu)$, and equation~\ref{eqn:Tmeasured1} to compute $\Delta T(\nu)$, our modelled values of $T_{\rm{refl-Gal}}(\nu)$ from Section~\ref{sec:reflected}, $T_{\rm{CMB}}=2.725$~K \citep{mather1994} and $T_{\rm{refl-Earth}}(\nu)=0~K$, since earthshine has been removed (see Section~\ref{sec:modelling}). As discussed in Section~\ref{sec:theory}, $T_{\rm{EoR}}(\nu)$ is ignored in this work. We then fit a power law with an offset to the data of the form:
\begin{equation} 
T_{\rm{Gal}}(\nu) - T_{\rm{moon}} = T_{\rm{Gal}150}\left(\frac{\nu}{150~\rm{MHz}}\right)^{\alpha} - T_{\rm{offset}},
\label{eqn:tgal_fit}
\end{equation}
where $\nu$ is the frequency in MHz and $T_{\rm{Gal}150}$, $\alpha$ and $T_{\rm{offset}}$ are the fitted parameters. Assuming that $T_{\rm{Gal}}(\nu)$ is equal to the power-law component, then $T_{\rm{moon}} = T_{\rm{offset}}$. We find the best-fit values for $T_{\rm{Gal}150}$, $\alpha$ and $T_{\rm{offset}}$ are $195\pm14$~K, $-2.64\pm0.14$ and $180\pm12$~K, respectively, where the quoted uncertainties are the $1\sigma$ errors derived from the diagonal elements of the covariance matrix of the fit. It must be noted, however, that the parameters are highly correlated, as shown by the near-unity values of the off-diagonal elements of the correlation matrix, $C$:\\
$C =
\left( \begin{array}{cccc}
  & T_{\rm{offset}} & T_{\rm{Gal}150} & \alpha  \\
T_{\rm{offset}} & 1.000 & 0.980 & 0.928 \\
T_{\rm{Gal}150} & 0.980 & 1.000 & 0.927 \\
 \alpha         & 0.928 & 0.927 & 1.000 \\
\end{array} \right)$. \\
In future work we will attempt to break this degeneracy, possibly by using an independent method to measure $T_{\rm{moon}}$. In fact, on- and off-Moon observations have already been taken with the Engineering Development Array \citep{EDA}, co-located with the MWA, that may suit this purpose.

\begin{figure*}
\centering 
\includegraphics[clip,trim=0 10 0 0,width=1.0\textwidth,angle=0]{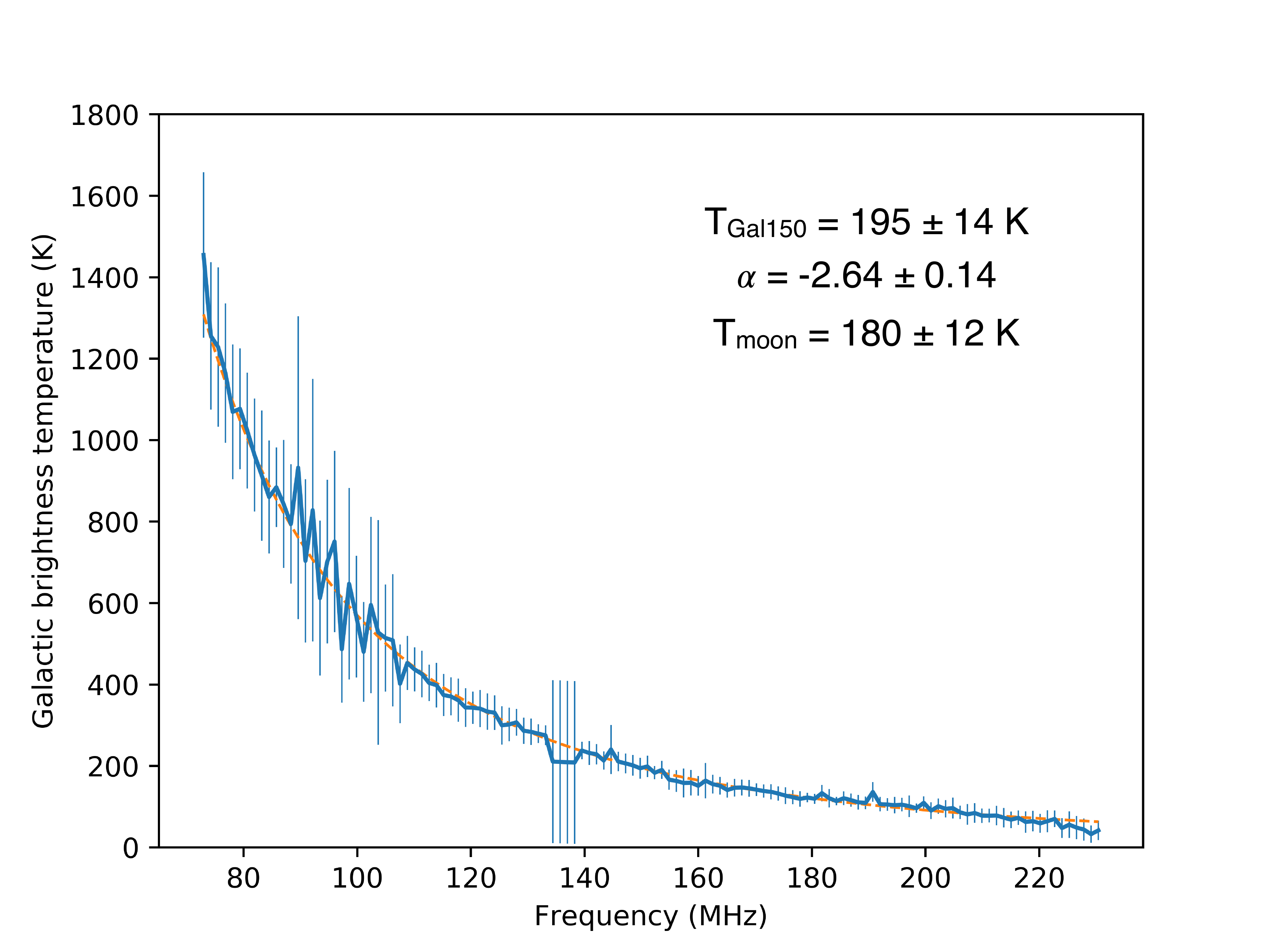}
\caption{The Galactic sky temperature, $T_{\rm{Gal}}(\nu)$, at the position of the Moon as measured with the MWA using the lunar occultation technique. The dashed orange line is a power-law fit to the data. A colour version of this figure is available in the online article.}
\label{inferred_temp}
\end{figure*}

In Fig.~\ref{inferred_temp} we plot:
\begin{equation} 
\begin{split}
T_{\rm{Gal}}(\nu) & = T_{\rm{moon}}+[T_{\rm{refl-Earth}}(\nu) + T_{\rm{refl-Gal}}(\nu)] \\ & \qquad \qquad - [\Delta T(\nu) + T_{\rm{CMB}}(\nu)],
\end{split}
\label{eqn:tgal_plot}
\end{equation}
where all variables are as described previously and $T_{\rm{moon}}~=~180$~K, from our fit above. The error bars are computed by propagation of the errors on $S_{m}(\nu)$, as shown in the bottom panel of Fig.~\ref{rfi_diffuse}. In Fig.~\ref{inferred_temp} we also plot our fit for $T_{\rm{Gal}}(\nu)$, described by equation~\ref{eqn:tgal_fit}.

\subsection{Global sky models}
\label{sec:global_sky_models}

In order to validate our results, we calculated the values we expect to obtain for $T_{\rm{Gal}}(\nu)$ using three available `global' sky models; the original GSM of \citet{gsm}, the updated GSM (hereafter GSM2017) of \citet{zheng2017}, and the Low Frequency Sky Model (LFSM) of \citet{dowell2017}. We generated sky maps for the GSM and GSM2017 using the \sc{PyGSM} \rm code\footnote{https://github.com/telegraphic/PyGSM} and for the LFSM we used the code provided on the Long Wavelength Array Low Frequency Survey website.\footnote{https://lda10g.alliance.unm.edu/LWA1LowFrequencySkySurvey/} Full sky maps were generated for each frequency band in our observations. 

We then calculated the mean temperature of the sky over the area occulted by the Moon in our observations, at sky position: RA~(J2000)~22\textsuperscript{h}48\textsuperscript{m}24.3\textsuperscript{s}, Dec~(J2000)~$-5$\degr 21$'$28.1$''$ (the Moon's position at UTC 2015 September 26 14:00:00, approximately halfway through the observations) at each frequency and for all three sky models. The corresponding Galactic latitude of the Moon in our observations is $-53.5$\degr.

\begin{figure*}
\centering 
\includegraphics[clip,trim=0 0 0 0,width=1.0\textwidth,angle=0]{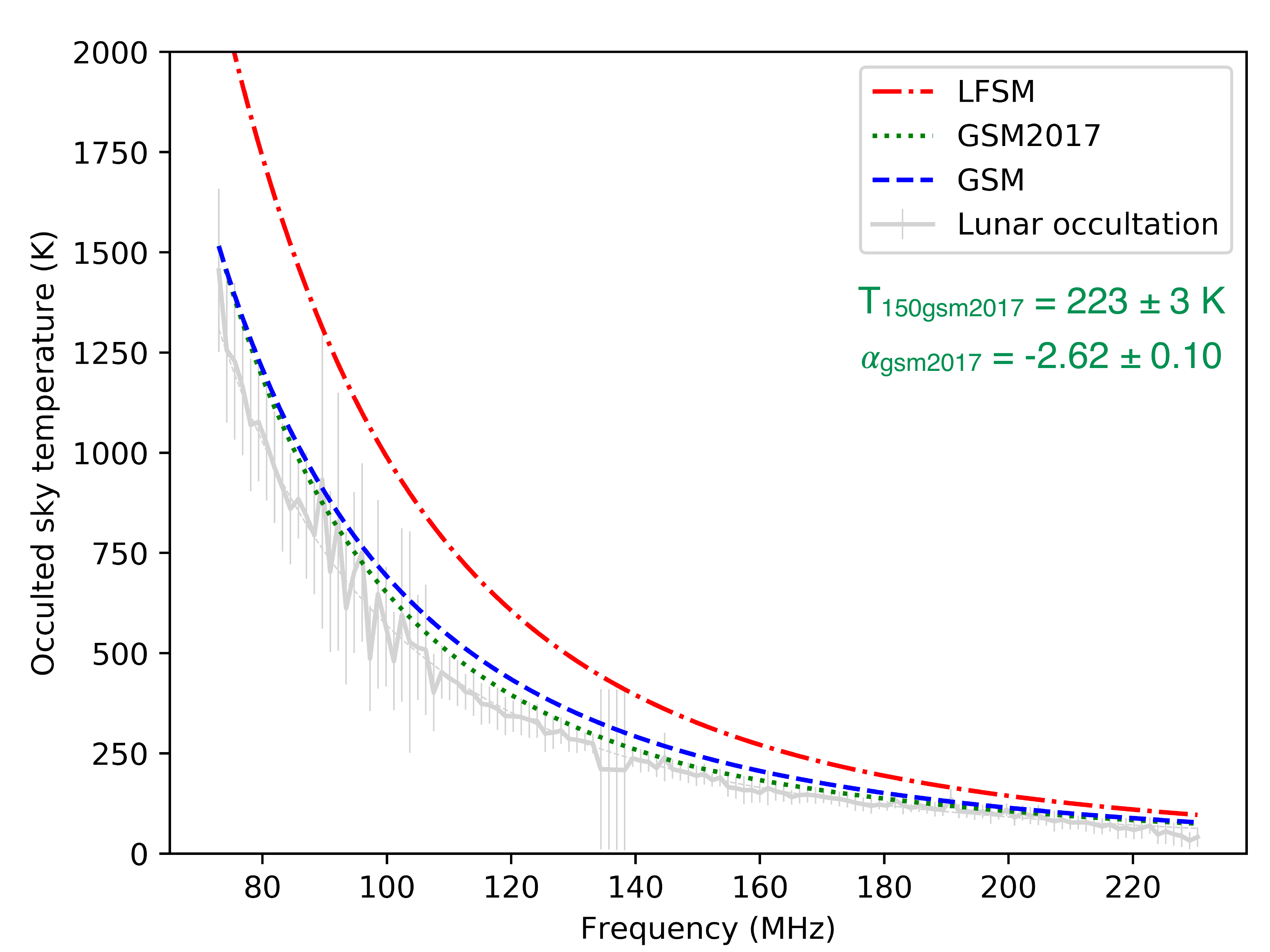}
\caption{Global sky model predictions for the mean brightness temperature of the occulted patch of sky using the GSM \citep{gsm}), the GSM2017 \citep{zheng2017} and the LFSM \citep{dowell2017}. The broken lines are fits to the data generated from each model assuming an error of 15\% on each point and using a simple power-law model. The unbroken line with error bars is our measured Galactic sky temperature reproduced from Fig.~\ref{inferred_temp} for comparison. A colour version of this figure is available in the online article.}
\label{fig:sky_models}
\end{figure*}

Assuming an error of 15\% on the occulted-sky average temperature data points generated from the global sky models, we fit a power law of the same form as equation~\ref{eqn:reflected_gal} to the data for each model. The resulting fits for the three global sky models are shown in Fig.~\ref{fig:sky_models}, along with our measurements of $T_{\rm{Gal}}(\nu)$ obtained using the lunar occultation technique, for comparison. For the GSM we find $T_{150}=242\pm3$ K and $\alpha=-2.59\pm0.10$, for the GSM2017 we find $T_{150}=223\pm3$ K and $\alpha=-2.62\pm0.10$ and for the LFSM we find $T_{150}=321\pm5$ K and $\alpha=-2.75\pm0.10$.

\section{discussion}
\label{sec:discussion}
\subsection{Measurement of the Galactic spectral index}
The most precise measurement of the Galactic synchrotron foreground in our observed radio-frequency range is that of \citet{mozdzen2017} using EDGES. At high Galactic latitudes matching our experiment they find a Galactic spectral index of $\alpha=-2.62\pm0.02$. Our result for the Galactic spectral index of $-2.64\pm0.14$ is consistent with the EDGES result (and the previous EDGES result of \citealt{rogers2008}), given the large uncertainties in our measurements. However it is not a clean comparison, since the area of sky covered by the EDGES beam is very large, covering a significant fraction of the entire sky, compared to our local measurement of the sky occulted by the Moon.

The best comparison to make is with respect to the three global sky models used to predict the measured sky temperature in Section~\ref{sec:global_sky_models}. We find that our spectral-index value of $\alpha=-2.64\pm0.14$ is consistent with the best-fit spectral index for the GSM model of $\alpha=-2.59\pm0.10$ and for the GSM2017 of $\alpha=-2.62\pm0.10$ (see Fig.~\ref{fig:sky_models}). The LFSM, however, greatly over-predicts the steepness of the spectral index and the temperature at 150~MHz. Our results are also consistent with the spectral index value (computed between 45~MHz and 408~MHz) of \citet{guzman2011}, who combine their own data with the map of \citet{haslam} to produce a spectral-index map that shows a value of around $-2.55$ at the position of our Moon observations.

In the only other experiment to use the lunar occultation technique, \citet{vedantham2015} found that a steeper spectral index of $\alpha=-2.9$ was the best fit to their data, at a lower frequency range of between 35 and 80~MHz with LOFAR. This steep spectral index is unexpected, particularly because their observations were at a low Galactic latitude of around $-10$\degr, where the Galactic synchrotron emission is expected to have a flatter spectral index of around $-2.4$ \citep{gsm,vedantham2015}. This could possibly be due to unaccounted-for diffuse earthshine, but the uncertainties in the measurements of \citet{vedantham2015} were also high due to problematic sidelobe sources in their data.

\subsection{Measurement of $T_{\rm{moon}}(\nu)$}
\label{sec:Tmoon}

\begin{figure*}
\centering 
\includegraphics[clip,trim=0 5 10 40,width=1.0\textwidth,angle=0]{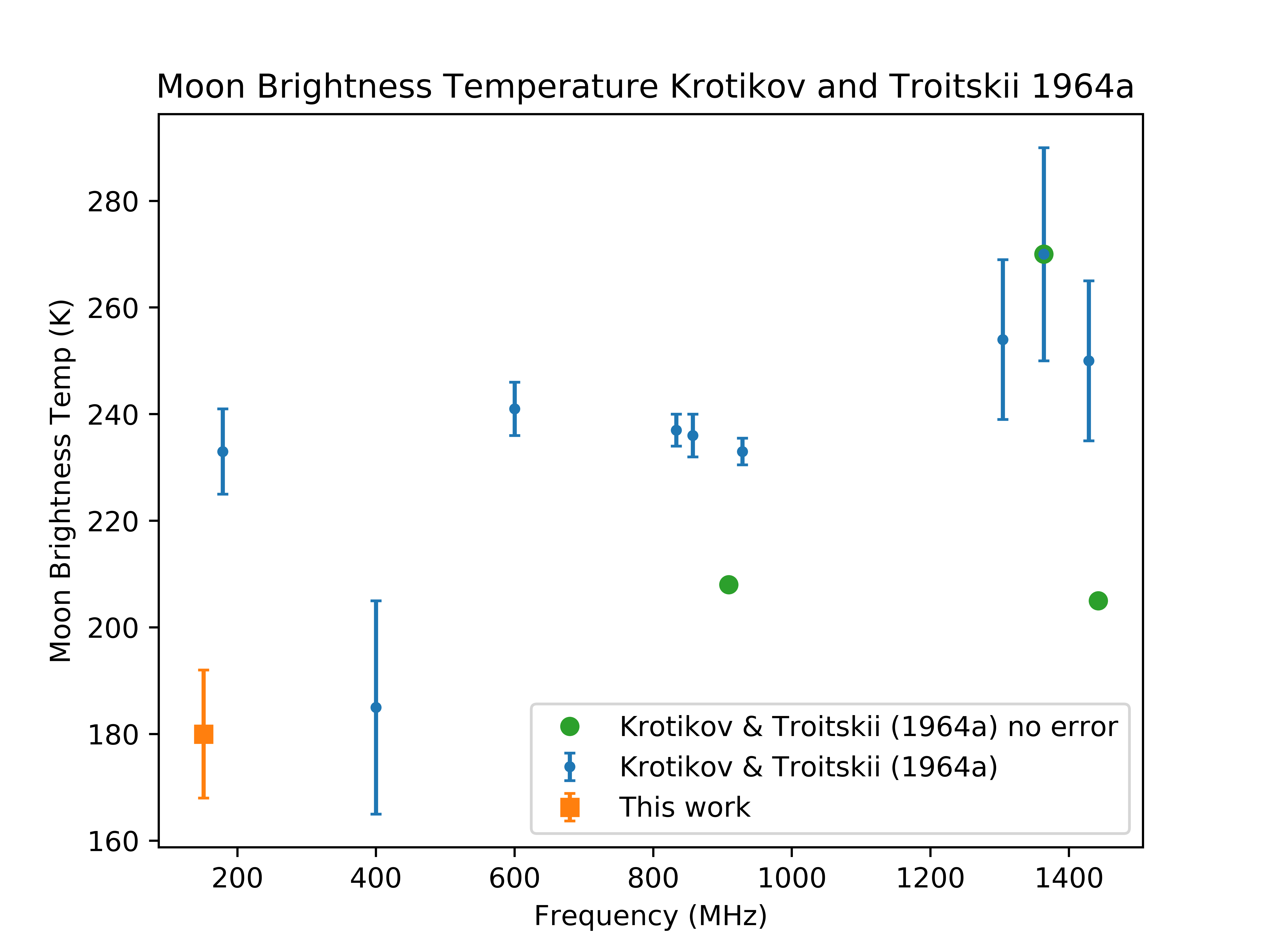}
\caption{The brightness temperature of the Moon from table 2 of \citet{krotikov1964a} and the brightness temperature of the Moon measured in this work. A colour version of this figure is available in the online article.}
\label{fig:Tmoon}
\end{figure*}

In a comprehensive review, \citet{krotikov1964a} report measured values of $T_{\rm{moon}}$ at wavelengths between 0.13~cm and 168~cm. In Fig.~\ref{fig:Tmoon} we plot our measurement of $T_{\rm{moon}}$ (with frequency taken to be in the middle of our band at 150~MHz), along with the values published in table~2 of \citet{krotikov1964a} that are below 1.5~GHz. Our value appears low compared to others, however it should be noted that there are few measurements at low frequencies (only one, \citealt{baldwin1961}, in our observing band) and that all of the measurements have large systematic errors, often not reflected in the quoted uncertainties. 

For example, the \cite{baldwin1961} measurement at 178~MHz makes use of an interferometer and therefore relies on an accurate knowledge of the mean sky temperature at the position of the Moon. The sky temperatures used are derived from those of \citet{turtle1962a} and \citet{turtle1962b}. Uncertainties of 10\%, plus a zero-level uncertainty of $\pm15$~K, are claimed by \citet{turtle1962a}, while \citet{turtle1962b}, who claim uncertainties in their overall sky temperature and zero-level calibration of a few K, use an antenna with a very large beamwidth ($15\degree$ in RA and $44\degree$ in DEC), so their measurements cannot be easily compared to the local temperature at the Moon's position in the sky.

Such systematic errors are present in many of the references quoted by \citet{krotikov1964a}, with most $T_{\rm{moon}}$ measurements relying on the sky temperature being known. This degeneracy between the mean sky temperature and the Moon temperature will also be problematic for new space-based experiments such as DARE and DSL, which aim to use the Moon as a shield to protect them from terrestrial RFI, as they seek to measure the global signal from the Dark Ages through to the EoR. New, low-frequency measurements of the Moon's brightness temperature such as ours, will be beneficial to these experiments as they plan and refine their calibration and imaging strategies.

It is also possible that the albedo of the Moon changes with frequency and that this is responsible for a change in the observed brightness temperature at low frequencies. A compilation of radar measurements presented by \citet{evans1969} shows that the total radar cross section of the Moon appears to remain relatively constant from cm to m wavelengths, but the error bars on the measurements are very large. This is despite the fact that varying the wavelength changes the apparent smoothness of the Moon. \citet{evans1969} suggest that the decrease in diffuse reflection with increasing frequency is offset by an increase in quasi-specular reflection. At lower frequencies, however, we are penetrating deeper into the lunar regolith \citep{krotikov1964a}, where the reflective properties are not well studied and changes in the Moon's albedo may be significant.

\subsection{RFI}
As with all global signal experiments, RFI is a problem as it is not smooth in frequency and could mimic an EoR signature. In our case, earthshine is the most significant form of RFI, especially in the FM radio band. However, using the techniques pioneered in this work, it may be that the bright FM radio band is of benefit, as it allows us to characterise the reflective properties of the Moon and remove both the diffuse and specular components of earthshine. Whether these improvements to the measured spectrum continue to reduce the error bars to a sufficient level as we include more data remains an open question.

\section{Conclusions and future work}
\label{sec:conclusion}

In this paper we have presented the first results from the MWA lunar occultation experiment. We have measured the Galactic synchrotron foreground and developed new techniques to mitigate against earthshine. We have also made a measurement of the brightness temperature of the Moon, which is of interest to low-frequency experiments proposed for lunar orbit.

Our initial results using the lunar occultation technique are promising. We are beginning to understand the errors and spectral features present in our data and will continue to refine our techniques. We will process the remaining data from the 2015 observing run using the techniques outlined in this paper. We also plan to investigate modelling techniques in $(u,v)$ space which may lead to improved fits and greater efficiency.

Observations covering a wide range of Galactic latitudes have been conducted during MWA observing semester 2018A, which provide a more diverse data set for the lunar occultation experiment. These observations utilise the long-baseline configuration of the MWA-Phase II (Wayth et al., in preparation), allowing us to map the quasi-specular earthshine at higher angular resolution. 

Future progress depends upon processing more data and further refining our techniques to effectively model foreground and reflected emission within our frequency range. The reflective behaviour of the Moon at low frequencies is not well studied and this will require particular attention. We must also develop techniques to break the degeneracy between the sky temperature and the Moon temperature in our fitting procedure. The prospects for measuring the redshifted 21-cm signal from the EoR and Cosmic Dawn using the lunar occultation technique with the MWA depend upon solving these key issues. We will report our progress in subsequent papers in this series.

\section*{Acknowledgments}

This scientific work makes use of the Murchison Radio-astronomy Observatory, operated by CSIRO. We acknowledge the Wajarri Yamatji people as the traditional owners of the Observatory site. Support for the operation of the MWA is provided by the Australian Government (NCRIS), under a contract to Curtin University administered by Astronomy Australia Limited. We acknowledge the Pawsey Supercomputing Centre, which is supported by the Western Australian and Australian Governments. BM is funded by a Discovery Early Career Researcher grant from the ARC (project number DE160100849). We acknowledge that the Bentley campus of the Curtin University of Technology, upon which most of this paper was written, is located on the land of the Noongar people. The Centre for All-Sky Astrophysics (CAASTRO) is an Australian Research Council Centre of Excellence, funded by grant CE110001020. Parts of this research were supported by the Australian Research Council Centre of Excellence for All Sky Astrophysics in 3 Dimensions (ASTRO 3D), through project number CE170100013. CMT is supported under the Australian Research Council's Discovery Early Career Researcher funding scheme (project number DE140100316). GB acknowledges support from the Royal Society and the Newton Fund under grant NA150184. This work is based on the research supported in part by the National Research Foundation of South Africa (grant No. 103424).








%
%


\bsp	
\label{lastpage}
\end{document}